\documentclass[aps,prb,twocolumn,showpacs,superscriptaddress]{revtex4-1}

\usepackage{graphicx}
\usepackage{amsmath}
\usepackage{bm}
\usepackage{longtable}
\usepackage{xspace}
\usepackage{color}

\newcommand{\camo}{CaMn$_7$O$_{12}$}
\newcommand{\srmo}{SrMn$_7$O$_{12}$}
\newcommand{\cdmo}{CdMn$_7$O$_{12}$}
\newcommand{\pbmo}{PbMn$_7$O$_{12}$}
\newcommand{\amo}{$A$Mn$_7$O$_{12}$}
\newcommand{\gmo}{$AA'_3B_4$O$_{12}$}

\begin{document}

\title{Universal magneto-orbital ordering in the divalent $A$-site quadruple perovskite manganites $A$Mn$_7$O$_{12}$ ($A$ = Ca, Sr, Cd, and Pb)}

\author{R. D. Johnson}
\email{roger.johnson@physics.ox.ac.uk}
\affiliation{Clarendon Laboratory, Department of Physics, University of Oxford, Oxford, OX1 3PU, United Kingdom}
\author{D. D. Khalyavin}
\author{P. Manuel}
\affiliation{ISIS facility, Rutherford Appleton Laboratory-STFC, Chilton, Didcot, OX11 0QX, United Kingdom}
\author{P. G. Radaelli}
\affiliation{Clarendon Laboratory, Department of Physics, University of Oxford, Oxford, OX1 3PU, United Kingdom}
\author{I. S. Glazkova}
\affiliation{Department of Chemistry, Lomonosov Moscow State University, Leninskie Gory, 119992 Moscow, Russia}
\author{N. Terada}
\affiliation{National Institute for Materials Science (NIMS), Sengen 1-2-1, Tsukuba, Ibaraki 305-0047, Japan}
\author{A. A. Belik}
\affiliation{Research Center for Functional Materials, National Institute for Materials Science (NIMS), 1-1 Namiki, Tsukuba, Ibaraki 305-0044, Japan}

\date{\today}

\begin{abstract}
Through analysis of variable temperature neutron powder diffraction data, we present solutions for the magnetic structures of \srmo, \cdmo, and \pbmo\ in all long-range ordered phases. The three compounds were found to have magnetic structures analogous to that reported for \camo. They all feature a higher temperature lock-in phase with \emph{commensurate} magneto-orbital coupling, and a delocked, multi-\textbf{k} magnetic ground state where \emph{incommensurate} magneto-orbital coupling gives rise to a constant-moment magnetic helix with modulated spin helicity. \cdmo\ represents a special case in which the orbital modulation is commensurate with the crystal lattice and involves stacking of fully and partially polarized orbital states. Our results provide a robust confirmation of the phenomenological model for magneto-orbital coupling previously presented for \camo. Furthermore, we show that the model is universal to the $A^{2+}$ quadruple perovskite manganites synthesised to date, and that it is tunable by selection of the $A$-site ionic radius.
\end{abstract}


\maketitle

\section{Introduction}

The quadruple perovskite manganites of general formula \gmo\ ($A$ = Na$^{+}$, Mn$^{2+}$, Ca$^{2+}$, Cd$^{2+}$, Sr$^{2+}$, Pb$^{2+}$, La$^{3+}$, Pr$^{3+}$, Nd$^{3+}$, or Bi$^{3+}$\cite{Ovsyannikov13,Marezio73,Johnson12,Bochu74,Glazkova15,Locherer12,
Prodi09,Mezzadri09_1,Mezzadri09_2,Imamura08}, $A'$ and $B$ = Mn), are host to a complex interplay between charge, orbital, and spin degrees of freedom \cite{Zeng99,Prodi04,Vasilev07,Gilioli14,Terada16,Johnson12}. The $A$ and $A'$ sites form a chemically ordered pseudo-cubic sublattice in the ratio $\tfrac{1}{4}$ $A$ to $\tfrac{3}{4}$ $A'$, accompanied by large, in-phase rotations of $B$-site oxygen octahedra ($a^+a^+a^+$ in Glazer notation), which also form a pseudo-cubic sublattice displaced from the $A$ and $A'$ sites along the pseudo-cubic body diagonal. In monovalent $A$-site NaMn$_7$O$_{12}$, checkerboard charge ordering of $B$-site Mn$^{3+}$ and Mn$^{4+}$ ions and CE-type orbital order and associated antiferromagnetism have been observed \cite{Prodi04,Streltsov14,Prodi14}, which are exactly equivalent to the long-range order found in the canonical manganite La$_{0.5}$Ca$_{0.5}$MnO$_3$ \cite{Radaelli97}. To the contrary, in divalent $A$-site \camo\ a 75\% Mn$^{3+}$ and 25\% Mn$^{4+}$ charge ordered phase (rhombohedral space group $R\bar{3}$) \cite{Przenioslo02} develops. This charge order is a crucial precursor to an unusual incommensurate magneto-orbital helix characterised by a multi-$k$ magnetic ground state \cite{Perks12,Cao15,Johnson16}, which is unlike any ordering phenomena found in the simple perovskite manganites. The trivalent $A$-site compounds LaMn$_7$O$_{12}$ and BiMn$_7$O$_{12}$ are similar to NaMn$_7$O$_{12}$, in that they display collinear, commensurate antiferromagnetic order, but in the absence of charge order \cite{Prodi09,Mezzadri09_2} (Mn$^{3+}$ only). It has therefore become apparent that the general type of long-range electronic ordering found in the \amo\ series can be selected by modification of the average valence of the $B$-site Mn ions through choice of the $A$-site cation charge state.

All divalent $A$-site quadruple perovskite manganites synthesised to date ($A$=Ca, Sr, Cd, and Pb) show a similar sequence of phase transitions. Below $T_\mathrm{OO}$, diffraction peaks have been observed \cite{slawinski2009,Perks12,Belik16_1} that originate in an orbital density wave propagating along the $c$-axis with wavevector $\mathbf{k_s}$ \cite{Perks12}. On further cooling, two magnetic phase transitions at $T_\mathrm{N1}$ and $T_\mathrm{N2}$ have been identified \cite{Zhang11,Glazkova15,Belik16_2}, with the exception of \pbmo\ for which an additional phase transition at an intermediate temperature, $T'$, was reported \cite{Belik16_2}. The magnetically ordered phases have only been well characterised for \camo \cite{Johnson12,Johnson16}. Below $T_\mathrm{N1}$ the manganese magnetic structure evolves continuously from a collinear spin density wave towards a constant-moment helix with modulated spin helicity, and the fundamental propagation vector $\mathbf{k_0}$ is locked into the periodicity of the orbital modulation with the relation $\mathbf{k_s} = 2\mathbf{k_0}$ (phase AFM1). At $T_\mathrm{N2}$ the fundamental magnetic order delocks from the orbital modulation periodicity in order to gain energy from spin exchange interactions, while the same magneto-orbital coupling persists through the creation of additional magnetic order parameters (phase AFM2). In AFM1, the magnetic order is \emph{commensurate} with the orbital order, and in AFM2 the magnetic order is \emph{incommensurate} with the orbital order. In both phases of \camo, the magneto-orbital textures are incommensurate with respect to the crystal lattice. Literature values for the transition temperatures of all four compounds are summarised in Table \ref{temptab}. $T_\mathrm{N1}$ was found to be remarkably similar across the series, however, significant variation was observed in $T_\mathrm{N2}$.

In this article we present results of neutron powder diffraction experiments, which show that the magneto-orbital phenomenology previously reported in \camo\ is universal to the $A^{2+}$ compounds studied. Furthermore, we show that the periodicity of the magnetic ground state, which is directly related to the variation in $T_\mathrm{N2}$, can be controlled by the $A$-site cation radius. The article is organised as follows. In Section \ref{exsec} we provide experimental details of sample growth, characterisation, and neutron powder diffraction. We present results on the commensurate orbital order in \cdmo\ (Sections \ref{CdOrbSec}), followed by results on the magnetic structures of \srmo, \cdmo, and \pbmo\ (Sections \ref{SrCaSec}, \ref{CdSec}, and \ref{PbSec}, respectively). The $A$Mn$_7$O$_{12}$ series is then discussed in general in Section \ref{dissec}. Finally, in Section \ref{consec} we summarise our conclusions.

\begin{table}
\caption{\label{temptab}Literature values for transition temperatures and the length of the orbital order propagation vector, $\mathbf{k_s}=(0,0,k_z)$, of the divalent $A$-site \amo\ compounds. Note that $T_\mathrm{N2}$ in \pbmo\ extends over a broad region of thermal hysteresis.}
\begin{ruledtabular}
\begin{tabular}{c | c c c c }
 & Cd\cite{Glazkova15,Belik16_1} & Ca\cite{slawinski2009,Slawinski10,Johnson12} & Sr\cite{Glazkova15,Belik16_1} & Pb\cite{Belik16_1,Belik16_2} \\
\hline
$r_\mathrm{ion}$ & 1.31 & 1.34 & 1.44 & 1.49 \\
\hline
$T_\mathrm{OO}$ [K] & 254 & 250 & 265 & 294 \\
$T_\mathrm{N1}$ [K] & 88 & 90 & 87 & 83 \\
$T'$ [K] & - & - & - & 77 \\
$T_\mathrm{N2}$ [K] & 33 & 48 & 63 & 37-65 \\
\hline
$|\mathbf{k_s}|$ & 2 & 2.0785 & 2.0765 & $\sim 2$
\end{tabular}
\end{ruledtabular}
\end{table}

\section{\label{exsec}Experiment}
Powder samples of \srmo, \cdmo, and \pbmo\ were prepared from stoichiometric mixtures of Mn$_2$O$_3$, MnO$_{1.839}$ (Alpha Aesar MnO2 99.997\% with the precise oxygen content determined by thermogravimetric analysis), PbO (99.999\%), CdO (99.99\%), and 4H-SrMnO$_3$. The mixtures were placed in Au capsules and treated at 6 GPa and 1373 K for 2 h for \cdmo\ and \pbmo\ and at 6 GPa and 1573 K for 2 h for \srmo\ (the duration of heating to the desired temperatures was 10 min) in a belt-type high-pressure apparatus. After the heat treatments, the samples were quenched to RT, and the pressure was slowly released. Data on \camo\ presented in this paper has been reproduced from previous studies \cite{Johnson12,Johnson16}, in which single crystals were grown at ambient pressure by the flux method \cite{Johnson12}, and then ground and sieved through a 35$\mu$m mesh. All samples were characterised by laboratory based x-ray powder diffraction, and heat capacity and magnetisation measurements using a Quantum Design PPMS and MPMS, respectively.

Neutron powder diffraction measurements were performed on the WISH time-of-flight diffractometer \cite{Chapon11} at ISIS, the UK Neutron and Muon Spallation Source. Samples were loaded into a cylindrical vanadium can and mounted within a $^4$He cryostat. Data were collected with high counting statistics at a fixed temperature within each magnetic phase, including paramagnetic for reference. Data were also collected with lower counting statistics on warming in the temperature range 1.5 K to 90 K. All diffraction data were refined using Fullprof \cite{Rodriguezcarvaja93}.

\section{Results}

\subsection{\label{AcSec}Average $R\bar{3}$ crystal structures}

\begin{figure*}
\includegraphics[width=18.0cm]{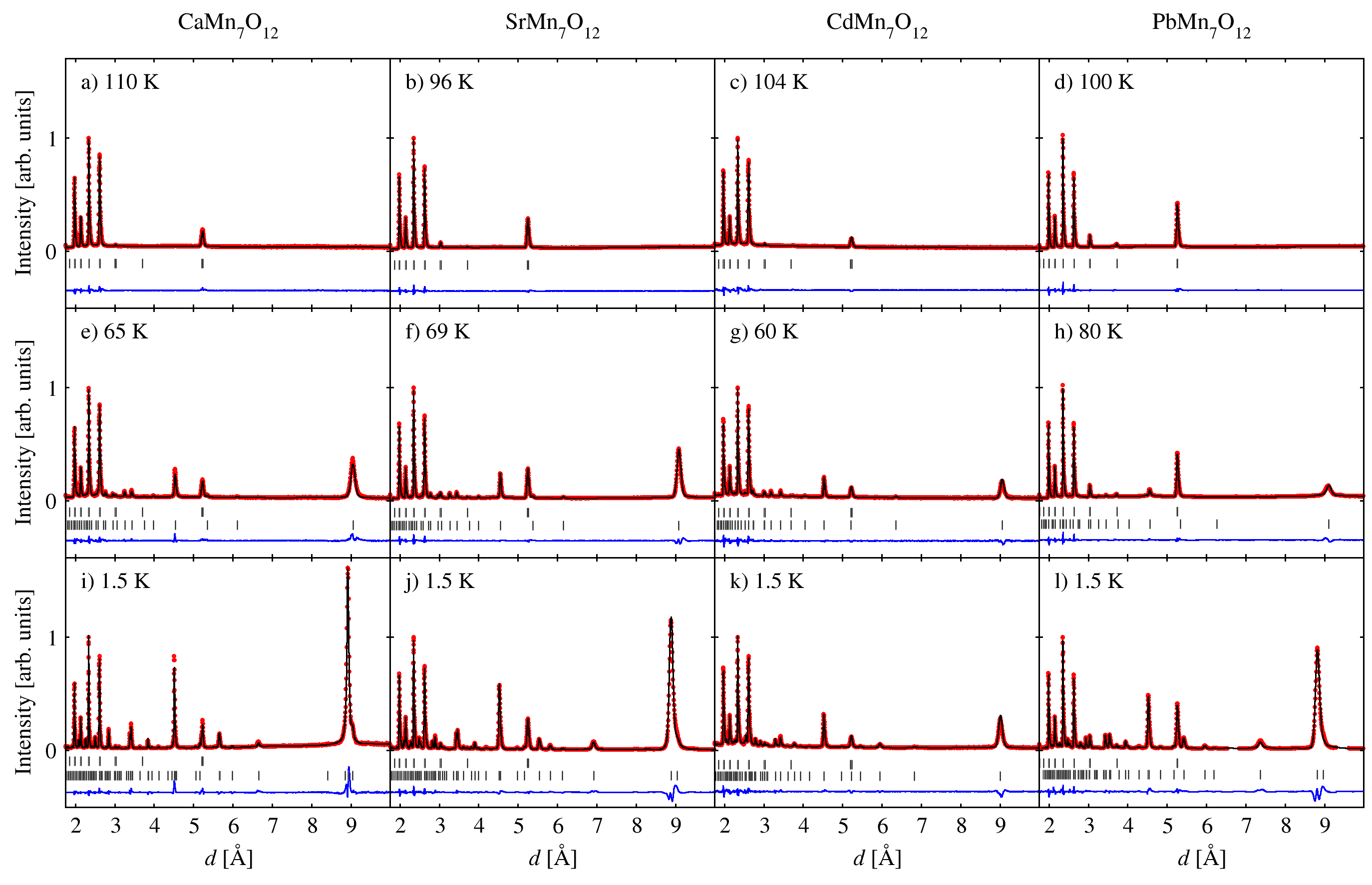}
\caption{\label{NPD_fig}Neutron powder diffraction data measured in bank 2 (average $2\theta = 58.3^\circ$) of the WISH diffractometer from \camo, \srmo, \cdmo, and \pbmo. Diffraction data (red points) collected in the paramagnetic, AFM1, and AFM2 phases are shown in a)-d), e)-h), and i)-l), respectively. The fitted nuclear (top tick marks) and magnetic (bottom tick marks) structural models are shown as a solid black line. A difference pattern ($I_\mathrm{obs}-I_\mathrm{calc}$) is given as a blue solid line at the bottom of each pane.}
\end{figure*}

Figure \ref{NPD_fig} shows neutron powder diffraction data measured in the paramagnetic, AFM1, and AFM2 phases of \srmo, \cdmo, and \pbmo, alongside data previously published on \camo \cite{Johnson12,Johnson16}. The diffraction patterns measured in the paramagnetic region $T_\mathrm{N1} < T < T_\mathrm{OO}$ (panes a-d in Figure \ref{NPD_fig}) were found to be qualitatively the same with the exception of weak superstructure reflections that originate in orbital modulations of differing periodicities. Direct comparisons were made by refining the \emph{average} rhombohedral crystal structure of \camo\ (space group $R\bar{3}$) against the four diffraction patterns. The respective fits (Figure \ref{NPD_fig}a-d) were in excellent agreement with the data, and the results are summarised in Table \ref{nucleartab}. Systematic variations in the lattice parameters and oxygen positions were observed due to the different ionic radii of the $A$-site cations.

Beyond the average $R\bar{3}$ structure, \camo, \srmo, and \pbmo\ support orbital order below $T_\mathrm{OO}$ that is incommensurate with respect to the crystal lattice (Table \ref{temptab}). The complex pattern of atomic displacements has been determined for \camo, in which they corresponded to a sinusoidal modulation of $d_{x^2-r^2}$ and $d_{y^2-r^2}$ orbital occupation of the single $B$-Mn$^{3+}$ $e_g$ electron. This incommensurate structure can be described within the four dimensional super-space group $R\bar{3}(00\gamma)0$ \cite{slawinski2009,Perks12}. To the contrary, in \cdmo\ the orbital order is commensurate with the crystal lattice (Table \ref{temptab}) \cite{Belik16_1}, such that the atomic displacements associated with the orbital order are in register with the average crystal structure. This material therefore presents a model case as it can be fully described within conventional 3D crystallography, as follows in Section \ref{CdOrbSec}.

\begin{table}
\caption{\label{nucleartab}Crystal structure parameters refined in the average $R\bar{3}$ space group for the paramagnetic phases.}
\begin{ruledtabular}
\begin{tabular}{c | c c c c }
 & \cdmo & \camo & \srmo & \pbmo \\
\hline
$r_\mathrm{ion}$ & 1.31 & 1.34 & 1.44 & 1.49 \\
$T$ [K] & 104 & 110 & 96 & 100 \\
\multicolumn{5}{l}{\textbf{Lattice Parameters} [$\mathrm{\AA}$]} \\
$a$  & 10.4394(2) & 10.4504(1) & 10.4866(1) & 10.5044(4)\\
$c$  & 6.3401(1) & 6.3472(1) & 6.3802(1) & 6.4088(3)\\
$V$  [$\mathrm{\AA}^3$] & 598.40(2) & 600.31(2) & 607.63(2) & 612.42(4)\\
\multicolumn{5}{l}{\textbf{Atomic fractional coordinates}\footnote{$A^{2+}$: [0,0,0], $A$-site Mn$^{3+}$: [0.5,0,0], $B$-site Mn$^{3+}$: [0.5,0,0.5], $B$-site Mn$^{4+}$: [0,0,0.5]}} \\
O1 $x$ &0.2274(4) & 0.2239(3) & 0.2260(3) & 0.2291(5)\\
$~\quad$ $y$ & 0.2779(4) & 0.2741(3) & 0.2755(3) & 0.2794(5)\\
$~\quad$ $z$ & 0.0826(5) & 0.0827(5) & 0.0817(4) & 0.0814(7)\\
O2 $x$ & 0.3420(4) & 0.3416(4) & 0.3412(4) & 0.3391(7)\\
$~\quad$ $y$ & 0.5226(4) & 0.5232(3) & 0.5213(3) & 0.5210(5)\\
$~\quad$ $z$ & 0.3440(9) & 0.3391(6) & 0.3360(5) & 0.3348(8)\\
\multicolumn{5}{l}{\textbf{Isotropic atomic displacement parameters, U$_\mathbf{iso}$} [$\mathrm{\AA}^2$]}\\
$A^{2+}$ & 0.037(3) & 0.031(3) & 0.016(2) & 0.030(2)\\
$A$-Mn$^{3+}$ & 0.0013(8) & 0.017(2) & 0.007(2) & 0.018(2)\\
$B$-Mn$^{3+}$ & 0.0013(8) & 0.010(2) & 0.004(2) & 0.012(3)\\
$B$-Mn$^{4+}$ & 0.0013(8) & 0.020(6) & 0.004(5) & 0.017(9)\\
O1 & 0.015(1) & 0.0121(11) & 0.001(1) & 0.002(1)\\
O2 & 0.017(1) & 0.0165(12) & 0.004(1) & 0.016(2)\\
\multicolumn{5}{l}{\textbf{Refinement reliability parameters} [\%]}\\
$R_{w}$ & 5.48 & 4.84 & 4.30 & 4.98\\
$R_\mathrm{Bragg}$ & 6.88 & 2.90 & 2.89 & 3.67\\
\end{tabular}
\end{ruledtabular}
\end{table}

\subsection{\label{CdOrbSec}\cdmo\ commensurate orbital order}

The commensurate nature of the orbital ordering in \cdmo\ can be explained by activation of the lock-in free-energy term, $\delta^3+\delta^{*3}$, which selects the $\mathbf{k_s}=(0,0,2)$ propagation vector (or equivalently $\mathbf{k_s}=(0,0,1)$). Here $\delta$ and $\delta^*$ are complex conjugate components of the structural order parameter associated with the orbital density wave, which transforms by the $\Lambda_1$ irreducible representation (see reference \citenum{Johnson16} for details). The presence of the commensurate phase does not exclude the existence of a hypothetical, high-temperature incommensurate orbital modulation similar to other members of the series, it only dictates that the periodicity of any incommensurate order would have to be sufficiently close to the commensurate order such that the lock-in term becomes competitive (and wins) at the measured temperatures. 

\begin{figure}
\includegraphics[width=8.8cm]{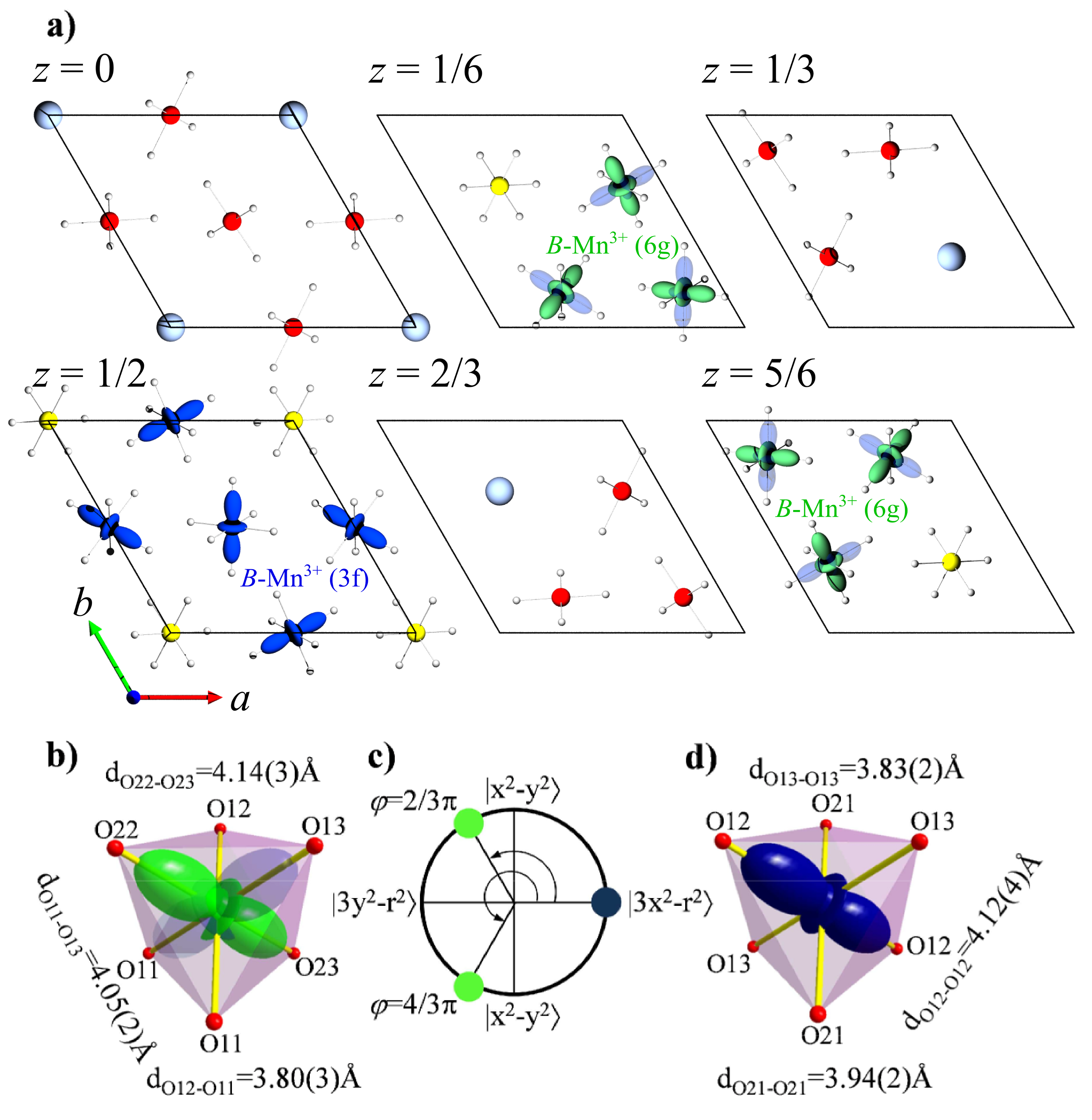}
\caption{\label{Cdorb}a) The $P\bar{3}$ crystal structure of \cdmo\ drawn projected down the hexagonal $c$-axis and in slices within a single unit cell. The $B$-site Mn$^{3+}$ Wyckoff position splitting in the $P\bar{3}$ subgroup and the respective commensurate orbital order is superimposed. $B$-site Mn$^{3+}$ orbitals are drawn in green and dark blue, $A$-site Cd$^{2+}$ ions in light blue, $A$-site Mn$^{3+}$ ions in red, $B$-site Mn$^{4+}$ ions in yellow, and oxygen in grey. The anisotropy of oxygen octahedra coordinating $B$-site Mn$^{3+}$ ions in the 6g and 3f Wyckoff positions, and the associated orbital states, are shown in (b) and (d), respectively. c) Diagram illustrating the relation between the orbital states of the $B$-site Mn$^{3+}$ ions and the phase of the commensurate structural modulation.}
\end{figure}

The commensurate propagation vector implies that the structural modulation on sites related by $R$-centring translations adopt phase shifts of $2/3\pi$ and $4/3\pi$. There are two isotropy subgroups associated with the $\Lambda_1$ irreducible representation and the $\mathbf{k_s}=(0,0,2)$ propagation vector, namely $P\bar{3}$ and $P3$ \cite{Campbell06,Stokes07}. These two cases correspond to different choices of the global phase of the modulation, which can be anticipated by inspecting the Wyckoff position splitting. The $P\bar{3}$ and $P3$ subgroups split the 9d ($B$-site Mn$^{3+}$) position into two (3f and 6g, see Figure \ref{Cdorb}) and three (3d, 3d and 3d) independent positions, respectively. The former implies that the orbital states of Mn$^{3+}$ in the sites related by the $R$-centring translations are identical (6g positions). In the latter case, Mn$^{3+}$ ions in these sites would be in different orbital states. Based upon the orbital modulation measured in \camo \cite{Perks12}, a $B$-site manganese orbital state, $\Psi$, can be defined by an admixture of $\psi_1 = |3x^2$-$r^2\rangle$ and $\psi_2 = |3y^2$-$r^2\rangle$ orbitals through the relation $\Psi = \psi_1|\cos (\phi/2)| + \psi_2|\sin (\phi/2)|$, as illustrated in Figure \ref{Cdorb}c. The $P\bar{3}$ symmetry sets $\phi$ to be 0 (or $\pi$) for the 3f position, and $\phi=2/3\pi$ or $\phi=4/3\pi$ for the sites in the 6g position, respectively (Figure \ref{Cdorb}c). Alternatively, the $P3$ subgroup implies an arbitrary phase, different from zero (or $\pi$), for all three independent 3d positions.

Our neutron diffraction data (Figure \ref{NPD_fig}c) was successfully refined using the $P\bar{3}$ subgroup, and the obtained structural parameters are summarised in Appendix \ref{appendix}. Using the non-centrosymmetric $P3$ symmetry did not improve the refinement quality. The choice of the global phase in the $P\bar{3}$ subgroup implies the presence of one third $B$-site Mn$^{3+}$ sites in the fully polarized orbital state (3f Wyckoff position, $\Psi = \psi_1$, see Figure \ref{Cdorb}). This fully polarised orbital state is unique to the $P\bar{3}$ structure, it is the most efficient configuration for releasing the Jahn-Teller instability, and it is therefore the likely microscopic mechanism behind the phenomenological lock in term $\delta^3+\delta^{*3}$. The other two third $B$-site Mn$^{3+}$ cations adopt the electronic configuration dominated by the $\psi_2$ orbital states ($\Psi = 1/2\psi_1 + \sqrt{3}/2\psi_2$).

The orbital phase, $\phi$, can be found for the measured Jahn-Teller anisotropy of the $B$-site Mn$^{3+}$ oxygen octahedra using the equation,
\begin{equation}
\tan(\tfrac{2}{3}\phi) = \frac{\sqrt{3}(l-s)}{(2m-l-s)}
\end{equation}
where $l$, $m$, and $s$ are the long, medium, and short bonds of the octahedra \cite{Wu11}, and the origin of $\phi$ is chosen such that it sweeps out the orbital sector $|3x^2$-$r^2\rangle$ to $|3x^2$-$r^2\rangle$. From the empirical bond lengths detailed in Figures \ref{Cdorb}b and \ref{Cdorb}d, $\phi \approx 0.18\pi$ and $\phi = 0.62\pi$ for the $3f$ and $6g$ sites, respectively. This result is in good agreement with the predicted orbital pattern quantified in Figure \ref{Cdorb}c, which demonstrates that the orbital modulation in \cdmo\ is indeed the commensurate limit of that found in \camo\cite{Perks12}.

\subsection{\label{SrCaSec}\srmo\ magnetic structures}

\begin{figure}
\includegraphics[width=8.8cm]{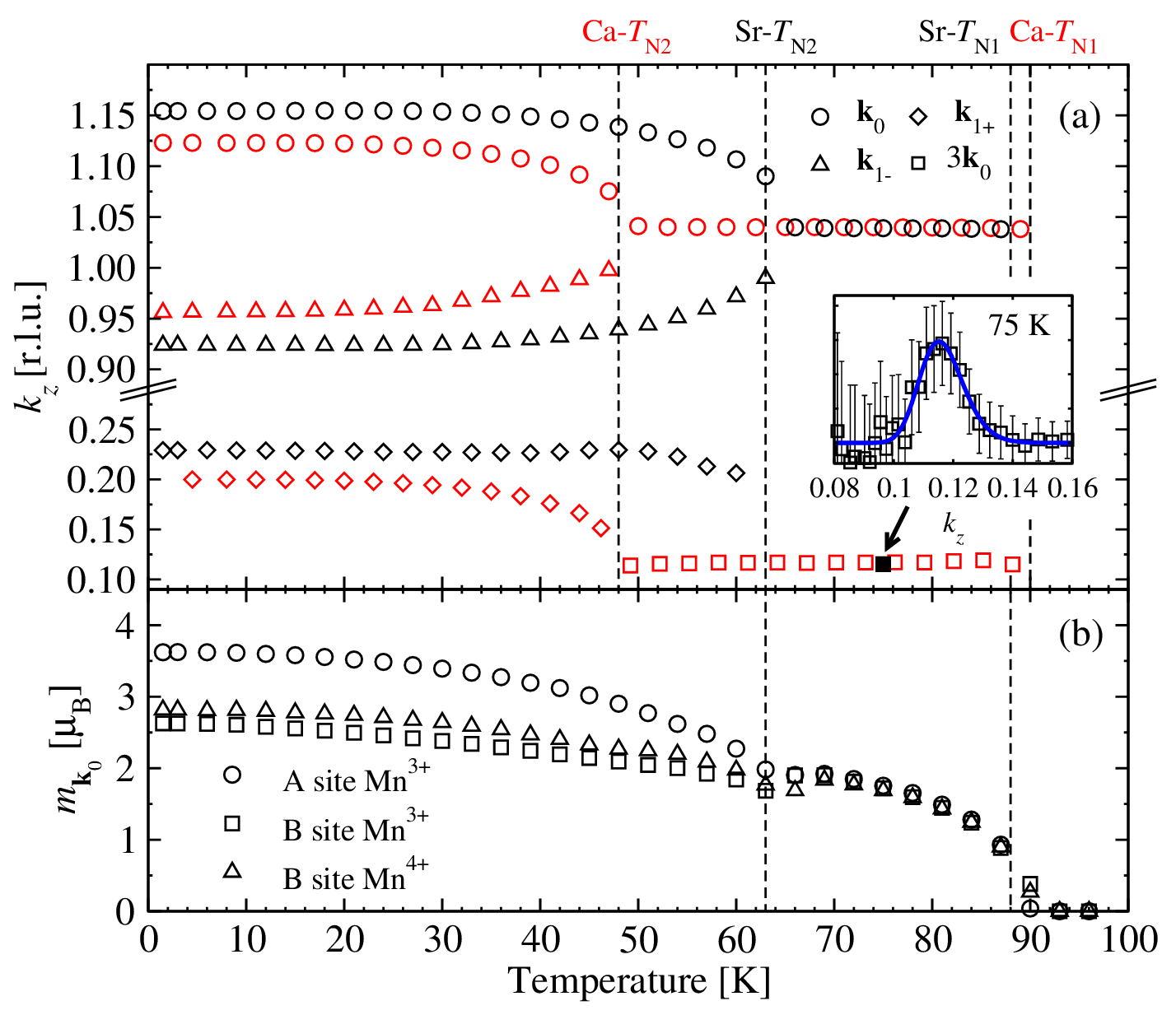}
\caption{\label{Srtempdep}a) Temperature dependence of the observed magnetic propagation vectors of \srmo\ (black symbols) measured on warming. The inset shows the third harmonic reflection measured at 75 K, and its position is marked on the main pane by the solid square. The temperature dependence of the magnetic propagation vectors measured from \camo\ are overlaid for comparison \cite{Johnson16} (red symbols). b) The temperature dependence of the \srmo\ moment amplitudes of the $\mathbf{k}_0$ component. In the main panes all error bars are smaller than the data points so they are omitted for clarity.}
\end{figure}

Immediately below $T_\mathrm{N1}$ (panes (e)-(f) in Figure \ref{NPD_fig}) additional diffraction peaks were observed that could be indexed with a fundamental magnetic propagation vector, $\mathbf{k_0}$, which lies parallel to the hexagonal $c$-axis. Both the incommensurate length of $\mathbf{k_0}$ in the AFM1 phase (Table \ref{ktab}), and the transition temperature, $T_\mathrm{N1}$, were found to be approximately equal for \srmo\ and \camo. The temperature dependence of $\mathbf{k_0}$ is given in Figure \ref{Srtempdep}a, which shows that in both compounds the fundamental modulation remains locked-in to the value $\tfrac{1}{2}|\mathbf{k_s}|$ down to $T_\mathrm{N2}$ (\emph{n.b.} $\mathbf{k_s}$ is also approximately equal for \srmo\ and \camo). Throughout AFM1, the fundamental modulation is accompanied by a third harmonic Fourier component with propagation vector $3\mathbf{k_0}$, as shown in the inset to Figure \ref{Srtempdep}a. A third harmonic was also observed in \camo, in which it arose through \emph{commensurate} magneto-orbital coupling of the form $\delta\eta_0\eta_3^* + \delta\xi_0\xi_3^* + \mathrm{c.c.}$, where c.c. indicates the complex conjugation, and $\delta$, $\eta_0$,$\xi_0$ and $\eta_3$,$\xi_3$ are components of the orbital ($\mathbf{k}_s$), fundamental magnetic ($\mathbf{k}_0$) and third harmonic magnetic ($3\mathbf{k}_0$) order parameters, respectively. We refer the reader to the Supplemental Information accompanying reference \citenum{Johnson16} for detailed definitions of the magnetic and orbital order parameters. This coupling results in a modulated spin helicity, \emph{i.e.} higher order Fourier components describe a periodic variation in the local spin helicity of an otherwise proper helix magnetic structure, in accordance with the variation in orbital occupation.

The helicity modulated AFM1 magnetic structure has nine parameters - three moment amplitudes of the symmetry inequivalent manganese sites, $A$-Mn$^{3+}$, $B$-Mn$^{3+}$, and $B$-Mn$^{4+}$, for each observed Fourier component (six amplitudes in total), two relative phases between the three sites, and a global phase. For an incommensurate structure the global phase can be arbitrarily set to zero, and the relative phase between $A$-Mn$^{3+}$ and $B$-Mn$^{3+}$ was fixed to a value of $\pi(|\mathbf{k}_0|-1)$ to be consistent with a helical magnetic structure. An insufficient number of third harmonic reflections were observed to reliably refine the respective amplitudes. Therefore, a model based upon just four parameters (three amplitudes and the $B$-Mn$^{4+}$ phase) was refined against the fundamental magnetic diffraction peaks measured in the AFM1 phase of \srmo. The fit showed excellent agreement with the data ($R_\mathrm{Bragg} = 2.8 \%$, $R_\mathrm{Mag} = 2.61\%$, Figure \ref{NPD_fig}f), the $B$-Mn$^{4+}$ phase refined to a value of 0.60(1), and the refined moment magnitudes are plotted in Figure \ref{Srtempdep}b.

\begin{table}
\caption{\label{ktab}Lengths of the magnetic propagation vectors of general form (0,0,$k_z$).}
\begin{ruledtabular}
\begin{tabular}{c | c c c c }
 & \cdmo & \camo & \srmo & \pbmo \\
\hline
$r_\mathrm{ion}$ & 1.31 & 1.34 & 1.44 & 1.49 \\
\multicolumn{5}{l}{\textbf{AFM1}}\\
$T$(K) & 60 & 65 & 69 & 80 \\
$|\mathbf{k_0}|$ & 1 & 1.03942(9) & 1.0391(1) & 1.0248(5)\\
\multicolumn{5}{l}{\textbf{AFM2} (1.5 K)}\\
$T$(K) & 1.5 & 1.5 & 1.5 & 1.5 \\
$|\mathbf{k_0}|$ & 1.0682(1) & 1.12354(8) & 1.15427(6) & 1.18410(8)\\
$|\mathbf{k_{1+}}|$ & 0.078(1) & 0.2031(4) & 0.2288(4) & 0.2376(5)\\
$|\mathbf{k_{1-}}|$ & 0.934(1) & 0.9554(4) & 0.9240(2) & 0.8717(4)\\
\end{tabular}
\end{ruledtabular}
\end{table}

The transition temperature $T_\mathrm{N2}$ was found to be 15 K higher when strontium was substituted for calcium. This difference sheds light on the nature of the phase transition, and is discussed in detail later. As shown in Figure \ref{Srtempdep}, $\mathbf{k}_0$ departs from lock-in below $T_\mathrm{N2}$, and monotonically increases towards its ground state length. The higher order modulation follows a similar trend, however, within AFM2 this order is no longer harmonic. Rather, it is related to the fundamental modulation through the sum $\mathbf{k}_{1+} = \mathbf{k}_s + \mathbf{k}_0$. Note that in the lock-in phase $\mathbf{k}_s = 2\mathbf{k}_0$, giving $\mathbf{k}_{1+} = 3\mathbf{k}_0$. As in \camo, the \srmo\ data showed an additional family of reflections that evolve below $T_\mathrm{N2}$ (Figure \ref{NPD_fig}j), which index with the propagation vector $\mathbf{k}_{1-} = \mathbf{k}_s - \mathbf{k}_0$. The modulations of the magnetic structure with propagation vectors $\mathbf{k}_{1+}$ and $\mathbf{k}_{1-}$, as found in AFM2, arise due to \emph{incommensurate} magneto-orbital coupling of the general form $\delta^n\eta_0\eta^*_{n+} + \delta^n\xi_0\xi^*_{n+} + \mathrm{c.c.}$ and 
$\delta^{*n}\eta_0\xi_{n-} + \delta^{*n}\xi_0\eta_{n-} + \mathrm{c.c.}$, where $n=1 \cdots \infty$ and the propagation vector of $\eta_{n\pm}$ and $\xi_{n\pm}$ is $\mathbf{k}_{n\pm} = n\mathbf{k}_s \pm \mathbf{k}_0$. Again, we refer the reader to the Supplemental Information accompanying reference \citenum{Johnson16} for details. The coupling invariants lead to the prediction of an infinite series of Fourier components, however, it was shown\cite{Johnson16} that for $n\geq 2$ the respective moment amplitudes approach the observation limit of our neutron powder diffraction experiment. This was indeed the case for \srmo.

Similar to the AFM1 magnetic structure model, the AFM2 model is parameterised in terms of three moment amplitudes per symmetry inequivalent manganese ion for each observed set of Fourier components with propagation vector $\mathbf{k}_{n\pm}$ (assuming that the magnetic structure has equal moment magnitudes on symmetry equivalent sites, the components with $\mathbf{k}_{n+}$ and $\mathbf{k}_{n-}$ must have equal moment amplitudes), plus three phases constrained as for AFM1. Rietveld refinement of this model against the \srmo\ data measured at 1.5 K gave an excellent fit to the data ($R_\mathrm{Bragg} = 2.77 \%$, $R_\mathrm{Mag} = 4.39\%$), as shown in Figure \ref{NPD_fig}j. The $B$-Mn$^{4+}$ phase refined to a value of 0.67(4), and the refined moment magnitudes are given in Table \ref{Amomtab}. The modulated helicity was found to reside predominantly on the $B$-Mn$^{3+}$ ions, and was negligibly small on the $B$-Mn$^{4+}$ ions. This is consistent with the fact that Mn$^{3+}$ is Jahn-Teller active and primarily responsible for the orbital order.

\begin{table}
\caption{\label{Amomtab}$A$Mn$_7$O$_{12}$ magnetic moment component magnitudes on each manganese sublattice, refined against data measured at 1.5 K.}
\begin{ruledtabular}
\begin{tabular}{c | c c c c }
 & \cdmo & \camo\cite{Johnson16} & \srmo & \pbmo \\
\hline
$r_\mathrm{ion}$ & 1.31 & 1.34 & 1.44 & 1.49 \\
\multicolumn{5}{l}{\bf{k$_0$ moments} ($\mu_\mathrm{B}$)} \\
$A$-Mn$^{3+}$ & 3.14(3) & 3.71(3) & 3.64(1) & 3.64(2)  \\
$B$-Mn$^{3+}$ & 2.24(3) & 2.69(2) & 2.58(2) & 2.34(2) \\
$B$-Mn$^{4+}$ & 2.67(6) & 3.09(6) & 2.92(3) & 2.94(3)  \\
\multicolumn{5}{l}{\bf{k$_{1+}$ \& k$_{1-}$ moments} ($\mu_\mathrm{B}$)} \\
$A$-Mn$^{3+}$ & 0.71(6) & 0.39(2) & 0.52(2) & 0.39(3)  \\
$B$-Mn$^{3+}$ & 1.33(6) & 1.41(2) & 1.47(2) & 1.52(2)  \\
$B$-Mn$^{4+}$ & 0.95(8) & 0.52(4) & 0.02(3) & 0.09(6)  \\
\end{tabular}
\end{ruledtabular}
\end{table}

In summary, \srmo\ displays the same magneto-orbital physics as observed in \camo. Below $T_\mathrm{N1}$ the magnetic and orbital order are both incommensurate with respect to the crystal structure, but their periodicities are locked together by commensurate magneto-orbital coupling. At $T_\mathrm{N2}$, which was found to be 15 K higher in \srmo, the magnetic and orbital periodicities de-lock, but remain incommensurately coupled down to 1.5 K. In both phases magneto-orbital interactions give rise to a modulation of the spin helicity, which is described by higher order Fourier components of the magnetic structure.

\subsection{\label{CdSec}\cdmo\ magnetic structures}
The magnetic structure of \cdmo\ in the AFM1 phase has been reported \cite{Guo17}, however, the AFM2 magnetic structure and coupling to the orbital physics had not been discussed. For completeness, we will begin with a repeat analysis of the AFM1 structure, and then continue with details of the AFM2 phase. We note that in the previous work \cite{Guo17} crystallographic phase separation was found below 60 K. We did not observe any such phase separation in the sample used in our neutron diffraction experiments, however, it was detected in another sample not reported here.

\begin{figure}
\includegraphics[width=8.8cm]{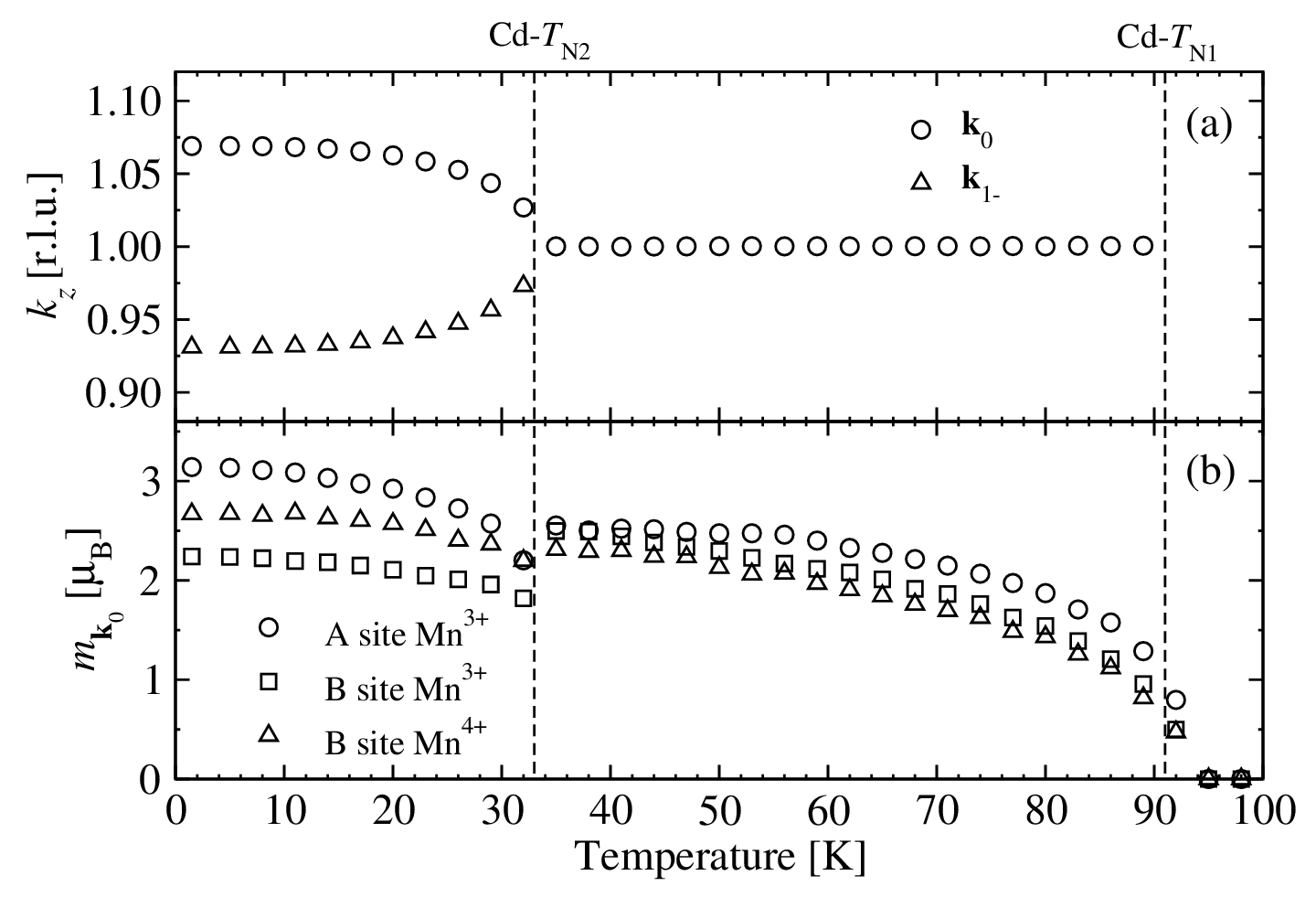}
\caption{\label{Cdtempdep}a) Temperature dependence of the observed magnetic propagation vectors of \cdmo\ measured on warming. b) The temperature dependence of the \cdmo\ moment amplitudes of the $\mathbf{k}_0$ component. All error bars are smaller than the data points so they are omitted for clarity.}
\end{figure}

When \cdmo\ orders magnetically below $T_\mathrm{N1}$, the magnetic diffraction peaks (Figure \ref{NPD_fig}g) index with the commensurate propagation vector $\mathbf{k}_0=(0,0,1)$, as previously observed \cite{Guo17}. Again, the periodicities of orbital and magnetic order within AFM1 are found to be locked together with the relationship $\mathbf{k}_s = 2\mathbf{k}_0$ (Figure  \ref{Cdtempdep}a). In this case the third harmonic cannot be identified as it overlaps exactly with much larger nuclear diffraction intensities with Miller indices $(0,0,l)$: $l=3m$, $m$ integer, \emph{i.e.}, in AFM1 the modulation of spin helicity occurs within a single conventional unit cell. 

In the absence of magneto-orbital coupling the AFM1 magnetic structure of \cdmo\ can be described by a single irreducible representation of space group $R\bar{3}$. In the commensurate limit manganese spins related by $R$-centring must be equal in magnitude and form a 120$^\circ$ configuration. However, as a result of the commensurate orbital ordering, each manganese sublattice is split into two sublattices, giving six, symmetry inequivalent Mn ions (see Appendix \ref{appendix}). One should then expect that, for example, the $B$-site Mn$^{3+}$ spins in the 3f and 6g Wyckoff positions do not form an exactly 120$^\circ$ configuration, nor should they have exactly the same magnitudes. The latter implies an admixture of a spin-density wave component, which is an essential condition for the lock in mechanism through magnetoelastic coupling \cite{Johnson16}. The parametrisation of the magnetic structure should be extended accordingly, however, it is the third harmonic modulation at $3\mathbf{k}_0=(0,0,0)$ that captures any departure from the $R\bar{3}$ AFM1 model - a natural consequence of the magneto-orbital coupling. As this component cannot be reliably refined, we continue with the $R\bar{3}$ AFM1 model as a good approximation.

\begin{figure}
\includegraphics[width=8.8cm]{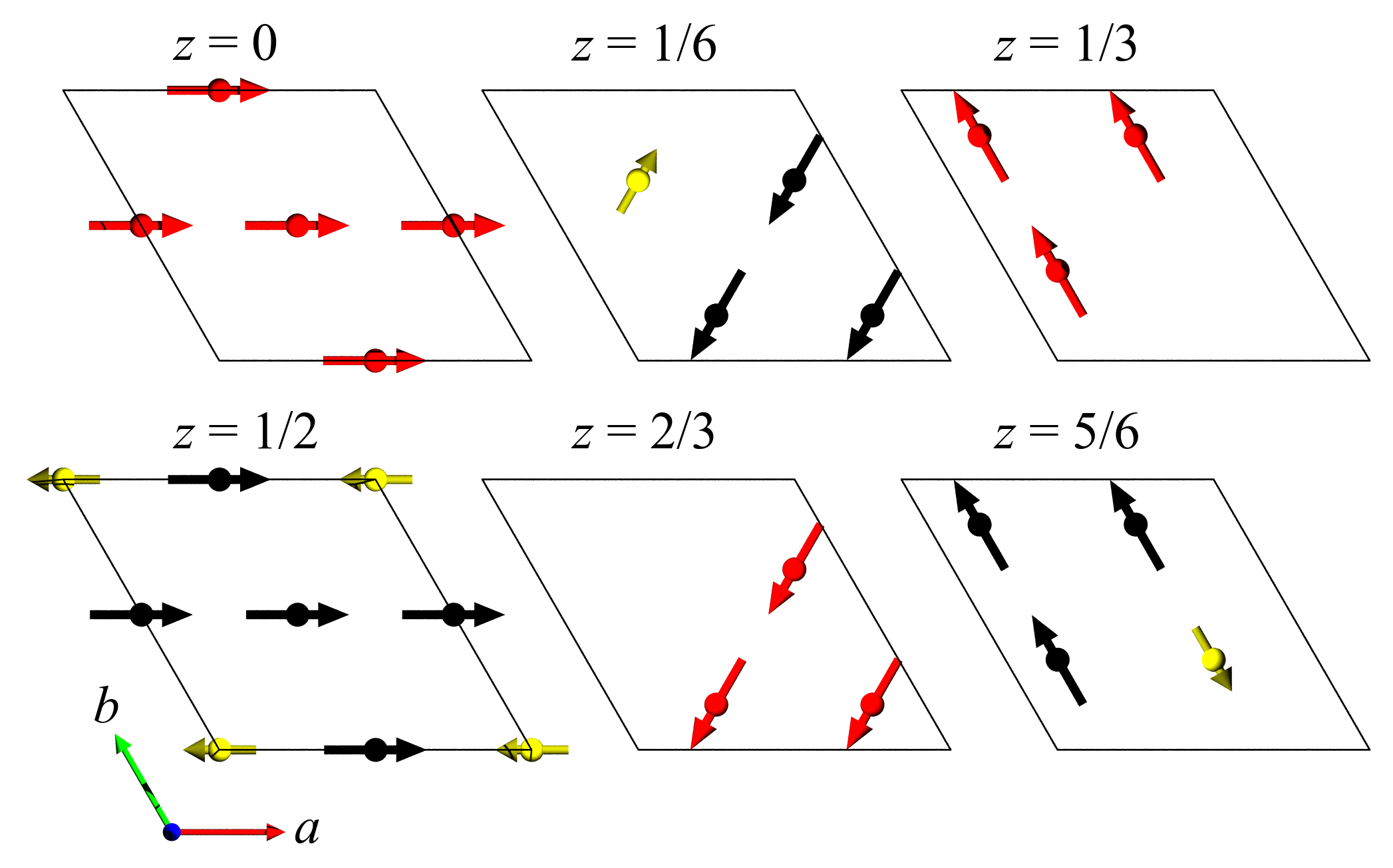}
\caption{\label{Cdmagstruc}The commensurate magnetic structure of \cdmo\ in the AFM1 phase, drawn projected down the hexagonal $c$-axis and in slices within a single unit cell. $A$-site Mn$^{3+}$, $B$-site Mn$^{3+}$, and $B$-site Mn$^{4+}$ magnetic moments are drawn as red, black, and yellow arrows, respectively. Note that the global moment directions have been drawn arbitrarily, with those at $z=0$ and $1/2$ set to be parallel to the $a$-axis. This direction cannot be determined by diffraction experiments, and it is independent of coupling to the orbital order.}
\end{figure}

A single-\textbf{k} AFM1 magnetic structure model was refined against the neutron powder diffraction data (Figure \ref{NPD_fig}g) with reliability factors of $R_\mathrm{Bragg} = 5.19 \%$ and $R_\mathrm{Mag} = 6.95\%$. The relative phases $A$-Mn$^{3+}$:$B$-Mn$^{3+}$, and $A$-Mn$^{3+}$:$B$-Mn$^{4+}$ were fixed to $\pi(|\mathbf{k}_0|-1)$ and freely refined to a value of 0.5(1), respectively. The refined moment magnitudes are plotted in Figure \ref{Cdtempdep}b, and the magnetic structure is illustrated in Figure \ref{Cdmagstruc}. Our results for the AFM1 phase are in good agreement with those previously published \cite{Guo17}.

Below $T_\mathrm{N2} = 33 \mathrm{K}$, the fundamental propagation vector de-locks from the orbital order propagation vector and monotonically increases towards the incommensurate ground state length (Figure \ref{Cdtempdep}a). The family of diffraction peaks with propagation vector $\mathbf{k}_{1-} = \mathbf{k}_s - \mathbf{k}_0$ also appears below $T_\mathrm{N2}$, as shown in Figure \ref{NPD_fig}k, which are consistent with the existence of incommensurate magneto-orbital coupling in the AFM2 phase. Although the $\mathbf{k}_{1+}$ intensities are no longer coincident with the nuclear intensities, they are not sufficiently separated to be observable beyond the experimental resolution. The AFM2 magnetic structure model was refined against the neutron powder diffraction data measured at 1.5 K (Figure \ref{NPD_fig}k) with reliability factors $R_\mathrm{Bragg} = 5.30 \%$ and $R_\mathrm{Mag} = 8.25\%$. The $B$-Mn$^{4+}$ phase refined to a value of 0.5(1), and the refined moment magnitudes are given in Table \ref{Amomtab}.

\subsection{\label{PbSec}\pbmo\ magnetic structures}

\begin{figure}
\includegraphics[width=8.8cm]{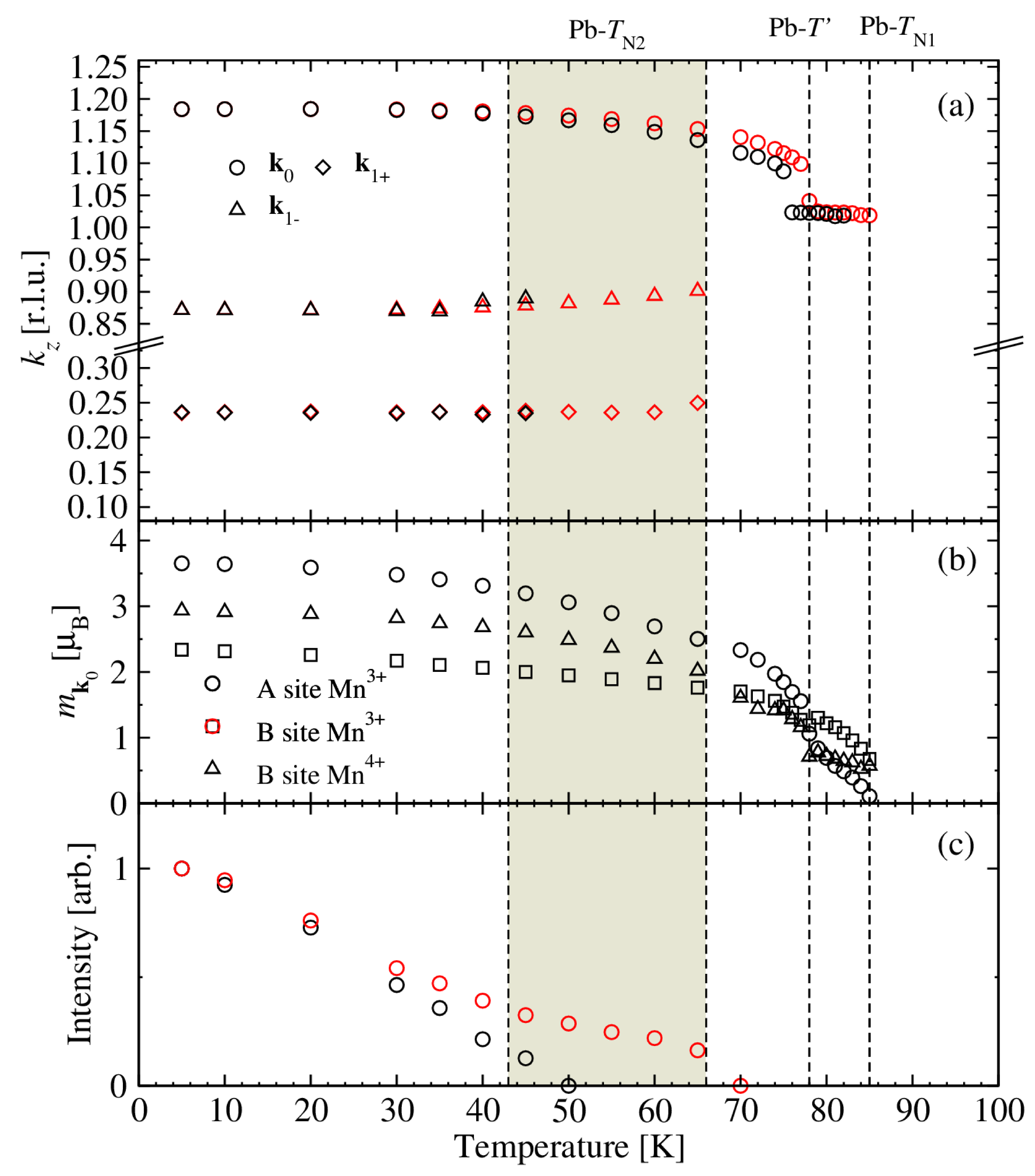}
\caption{\label{Pbtempdep}a) Temperature dependence of the observed magnetic propagation vectors of \pbmo\ measured on cooling (black symbols) and warming (red symbols). The large region of hysteresis is shaded.  b) The temperature dependence of the \pbmo\ moment amplitudes of the $\mathbf{k}_0$ component measured on warming. c) The temperature dependence of the integrated intensity of the \textbf{k}$_{1+}$ diffraction peak measured on cooling (black cirlces) and warming (red circles). All error bars are smaller than the data points so they are omitted for clarity.}
\end{figure}

\pbmo\ was found to magnetically order with the same AFM1 lock-in magnetic structure as observed in \camo\ and \srmo. The fundamental magnetic diffraction peaks, shown in Figure \ref{NPD_fig}h, could be indexed with the incommensurate propagation vector $\mathbf{k}_0 = (0,0,1.0248)$. The third harmonic diffraction intensity with Miller indices $(0,0,0.0744)$ corresponds to a d-spacing of 86.14 $\mathrm{\AA}$, which is just outside the practical measurement range of the WISH diffractometer (the same reflection for \camo\ and \srmo\ could be observed with d-spacing values of 53.67 $\mathrm{\AA}$ and 54.39 $\mathrm{\AA}$, respectively). We note that the observation of a third harmonic in the AFM1 phase provides the strongest evidence for a locked-in magneto-orbital helix. We therefore conclude that the AFM1 magnetic structure in \pbmo\ is a lock-in phase based upon the numerical relationship between the respective propagation vectors alone. The diffraction data collected in the AFM1 phase was fit with the single-\textbf{k} AFM1 model (Figure \ref{NPD_fig}h). Reliability factors of $R_\mathrm{Bragg} = 3.43\%$ and $R_\mathrm{Mag} = 6.48\%$ were obtained, the relative phase of the $B$-site Mn$^{4+}$ moments was fixed to $\pi$ to achieve convergence of the refinement, and the temperature dependence of the moment magnitudes are plotted in Figure \ref{Pbtempdep}.

The ground state magnetic structure of \pbmo\ was found to be delocked from the orbital order periodicity, and the observation of \textbf{k}$_{1+}$ and \textbf{k}$_{1-}$ magnetic diffraction peaks (Figure \ref{NPD_fig}l) indicated the presence of incommensurate magneto-orbital coupling and the same AFM2 magnetic structure as observed in the ground state of \camo, \srmo, and \cdmo. The AFM2 magnetic structure model was refined against the data giving reliability factors 
$R_\mathrm{Bragg} = 4.23 \%$ and $R_\mathrm{Mag} = 7.71\%$. The relative phase of the $B$-site Mn$^{4+}$ moments was again fixed to $\pi$ to achieve convergence of the refinement, and the moment magnitudes are given in Table \ref{Amomtab}.

Whilst the ground state and AFM1 magnetic structures were found to be analogous to those observed in the other AMn$_7$O$_{12}$ compounds, \pbmo\ displayed very different behaviour over an intermediate range of temperatures (Figure \ref{Pbtempdep}). Instead of supporting a single phase transition from the AFM1 phase to the AFM2 phase at $T_\mathrm{N2}$, the magnetic structure of \pbmo\ delocked on cooling below $T'\sim 78$ K, but orders with a single propagation vector, \textbf{k}$_{0}$, down to $T_\mathrm{N2} \sim 43$ K. Below $T_\mathrm{N2}$ the system entered the AFM2 phase, characterised by the appearance of \textbf{k}$_{1+}$ and \textbf{k}$_{1-}$ Fourier components, as shown in Figure \ref{Pbtempdep}c. On warming, a large region of hysteresis is observed whereby the system did not leave the AFM2 phase until $T_\mathrm{N2} \sim 66$ K, above which there exists the same delocked, single-\textbf{k} phase up to $T'$ (Figure \ref{Pbtempdep}). The presence of \textbf{k}$_{1+}$ and \textbf{k}$_{1-}$ magnetic diffraction peaks below the lock-in phase transition is a requirement of the magneto-orbital helical state. Hence, their absence in \pbmo\ in the intermediate temperature range $T_\mathrm{N2} < T < T'$ indicates that either the two electronic textures temporarily decouple, or the orbital modulation is temporarily destroyed. Establishing the true nature of the phase transition at $T'$ will require a detailed study of the crystal structure.

\section{\label{dissec}Discussion}

\begin{figure}
\includegraphics[width=8.8cm]{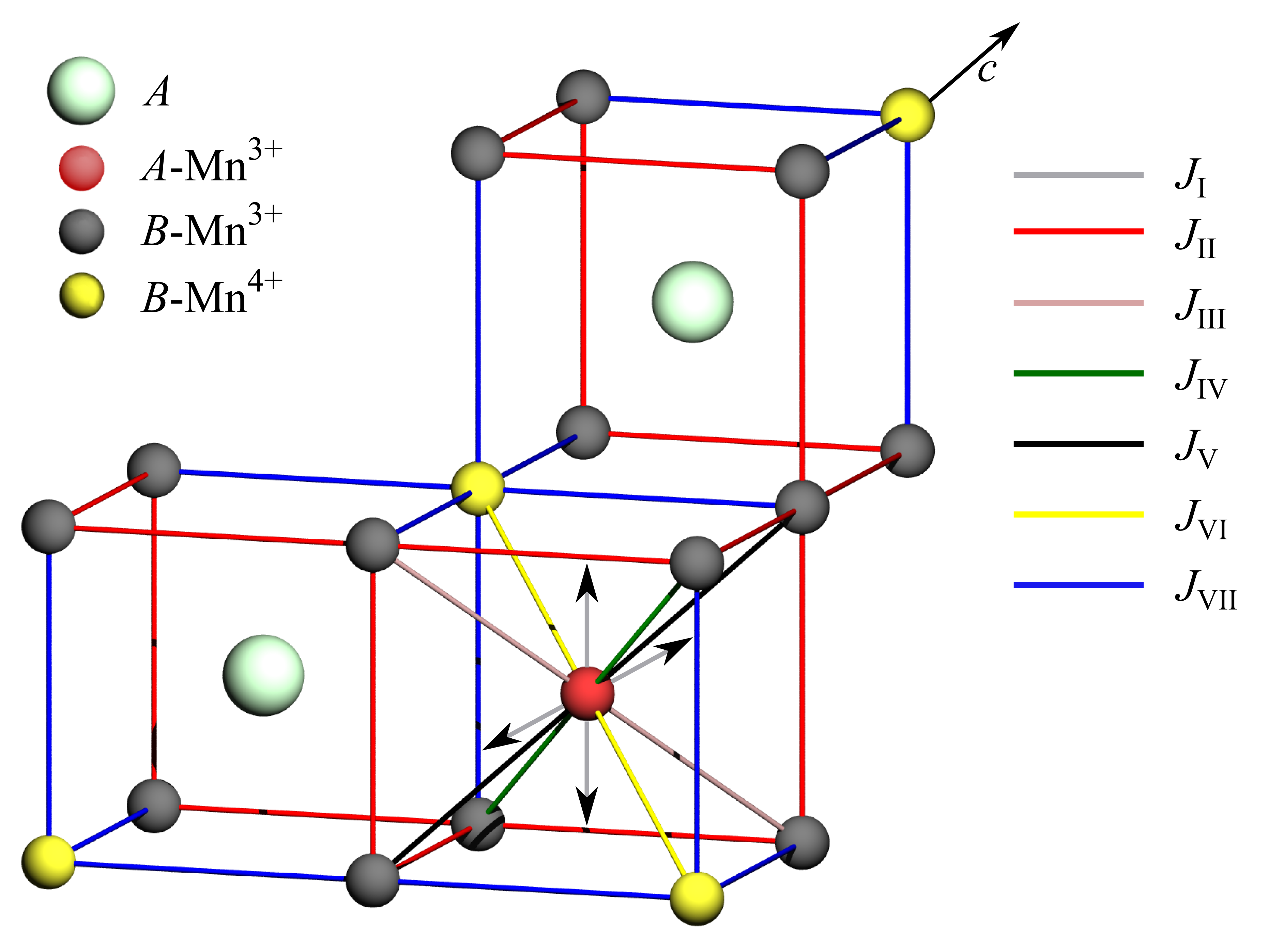}
\caption{\label{Js}The seven symmetry inequivalent nearest neighbour magnetic exchange interactions of the average ($R\bar{3}$) $A$Mn$_7$O$_{12}$ crystal structure. Two $B$-site pseudo-cubic sublattice units centred on the $A^{2+}$ cations are shown, which share faces with $B$-site pseudo-cubic units centred on the $A$-Mn$^{3+}$ sites. A single $A$-Mn$^{3+}$ cube is drawn for clarity, in which arrows indicate connectivity between neighbouring $A$-Mn$^{3+}$ ions. Note that $J_\mathrm{VII}$ and $J_\mathrm{II}$ are equivalent to $J_1$($J_2$) and $J_3$ in the model presented in reference \citenum{Perks12}.}
\end{figure}

As we have seen, the lock-in and ground state magneto-orbital coupling mechanisms\cite{Johnson16} and the resulting magnetic structures are in essence identical in \camo, \srmo, \cdmo, and \pbmo. The magneto-orbital textures differ only in their periodicity, which is determined in the ground state by the competing magnetic exchange interactions. In the remainder, we demonstrate that the exchange interactions that determine the ground-state propagation vector, $\mathbf{k}_0$, are effectively tuned by changing the ionic radius of the $A$-site cation.  Crucially, this can occur without changing the average antiferromagnetic interactions, and hence the Neel temperature $T_\mathrm{N1}$. We construct a mean field model of the magnetic exchange energy within the average $R\bar{3}$ $A$Mn$_7$O$_{12}$ crystal structure, which contains 7 symmetry inequivalent nearest-neighbour magnetic exchange interactions, as illustrated in Figure \ref{Js}. The exchange parameter space can be reduced as geometry allows two pairs of interactions, ($J_\mathrm{I}$ + $J_\mathrm{II}$) and ($J_\mathrm{III}$ + $J_\mathrm{IV}$), to be factorised, and we write the combined exchange interactions as $J_\mathrm{I+II}$ and $J_\mathrm{III+IV}$, respectively. The ground state Heisenberg mean field energy was calculated for the five dimensional parameter space, whilst keeping $J_\mathrm{V}$ fixed to -1 (ferromagnetic) - consistent with the experimentally determined magnetic structures. By imposing empirical constraints on the relative phases of the three manganese sublattices, one can show that $J_\mathrm{VI} \approx 0$, $J_\mathrm{I+II}$ and $J_\mathrm{III+IV} > 0$ (antiferromagnetic), and $J_\mathrm{VII} < 0$ (ferromagnetic). The exchange energy for $A$Mn$_7$O$_{12}$ can therefore be written in terms of just three interactions, $J_\mathrm{I+II}$, $J_\mathrm{III+IV}$, and $J_\mathrm{VII}$, relative to $J_\mathrm{V}$:
\begin{eqnarray}
\Delta E = (6J_\mathrm{I+II} - 3J_\mathrm{VII})\cos(\tfrac{2}{3}\pi k_z)  &-& 2J_\mathrm{III+IV}\cos(\tfrac{1}{3}\pi k_z) \nonumber \\
&-& J_\mathrm{V}\cos(\pi k_z)
\end{eqnarray}

\begin{figure}
\includegraphics[width=8.8cm]{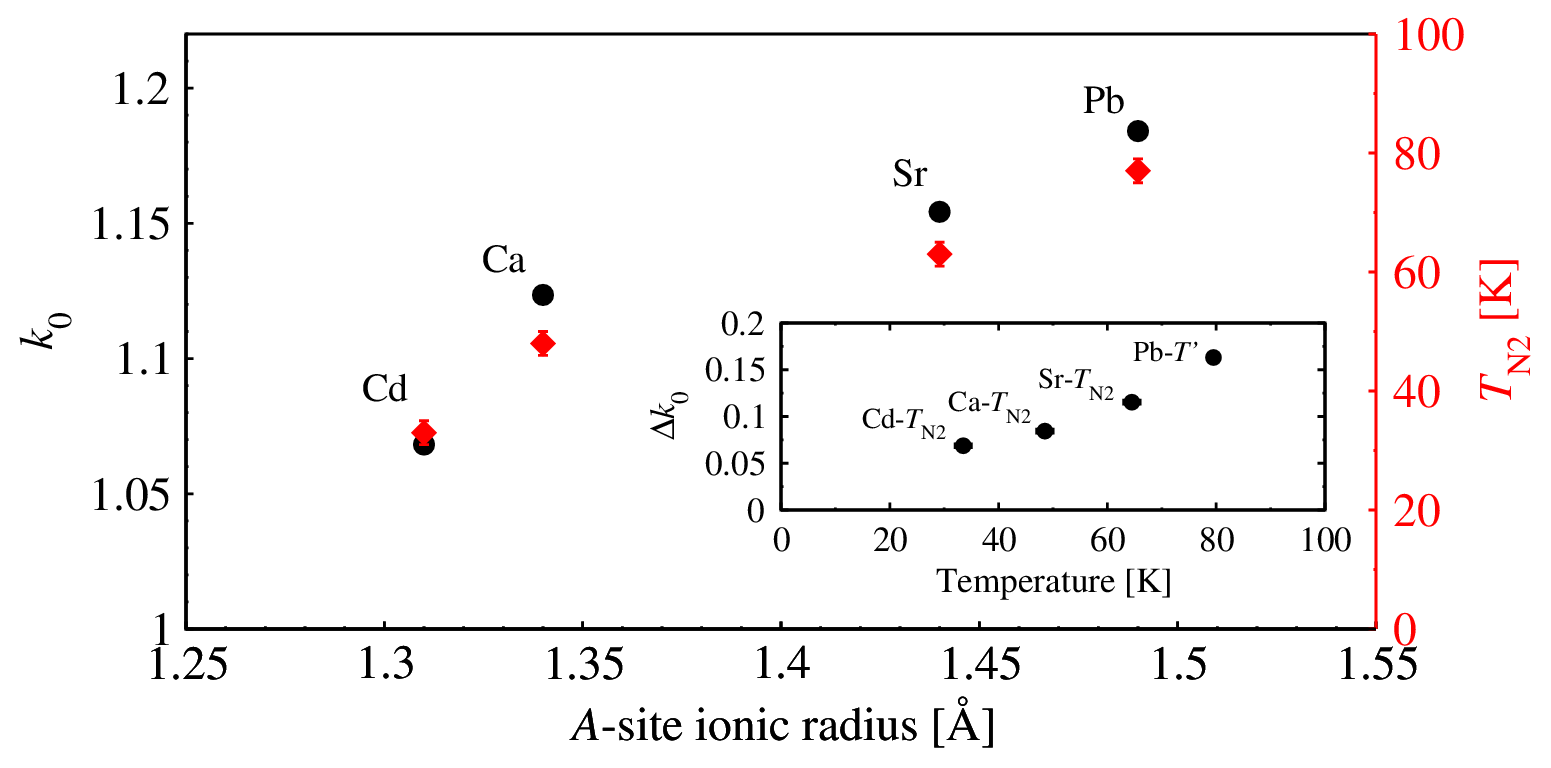}
\caption{\label{Acomp}The ground state fundamental propagation vector length, $k_0$, (black circles) and the lock-in transition temperature, $T_\mathrm{N2}$, (red diamonds) plotted against the $A$-site ionic radius for all measured compounds. The inset shows $\Delta k_0$, the difference between the ground state value and the AFM1 lock-in value of $k_0$, against $T_\mathrm{N2}$.}
\end{figure}

Minimising $\Delta E$ with respect to $k_z$ gives
\begin{equation}
k_z = \frac{3}{\pi}\tan^{-1}\Bigg(\sqrt{\frac{8}{2+\alpha^2+\beta-\alpha\sqrt{4+\alpha^2+2\beta}} - 1}\Bigg)
\end{equation}
where $\alpha = (4J_\mathrm{I+II}-2J_\mathrm{VII})$ and $\beta = \tfrac{4}{3}J_\mathrm{III+IV}$. For small $\beta$, the commensurate structure ($k_z = 1$) is approached as $\alpha$ tends to zero. The propagation vector varies rapidly for small, positive alpha, and then asymptotically approaches a value of 1.5. Accordingly, the ground state propagation vector in $A$Mn$_7$O$_{12}$ can be gradually increased by $J_\mathrm{I+II}$ becoming more positive, $J_\mathrm{VII}$ becoming more negative, or a uniform increase in the magnitude of both. Figure \ref{Js} shows the pseudo-cubic network of $J_\mathrm{II}$ and $J_\mathrm{VII}$ pathways of the $B$-site sublattice, centred on the $A$-site cation. By changing the radius of the central cation, the bonding geometry of the relevant exchange pathways will be modified resulting in a change in the ground state $k_z$, as found experimentally and summarised in Figure \ref{Acomp}.

Figure \ref{Acomp} also shows $T_\mathrm{N2}$ ($T'$ in \pbmo) plotted against the $A$-site cation radius, which shows a similar trend to the ground state $k_z$. At all temperatures below $T_\mathrm{N1}$ the free energy gain from magneto-orbital lock-in is in competition with energy lost from magnetic exchange due to the required deviation from the ground state magnetic propagation vector. The lock-in phase transition therefore arises at a temperature above which the entropy of the system stabilises locked-in magneto-orbital coupling. In this scenario, the lock-in transition temperature would be zero when the energy lost though exchange exactly equals the energy gain through lock-in. Then by monotonically increasing the difference in exchange energy between the ground state and the lock-in magnetic structure, the transition temperature becomes finite and also increases monotonically, \emph{i.e.}, the lock-in transition temperature is expected to be dependent on the propagation vector difference $\Delta k_0 = |\mathbf{k_0}| - |\mathbf{k_s}|/2$, where $|\mathbf{k_0}|$ is evaluated in the ground state. The inset to figure \ref{Acomp} shows $\Delta k_0$ plotted against the lock-in transition temperature for \cdmo, \camo, \srmo, and \pbmo, which clearly demonstrates the expected dependence. Whilst $T_\mathrm{N2}$ is mostly determined by the \emph{relative} magnitude of competing exchange interactions, at the mean field level $T_\mathrm{N1}$ is expected to be dependent upon the \emph{average} magnitude of the exchange interactions, and hence independent of the $A$-site cation radius, as observed.

\section{\label{consec}Conclusions}
The magnetic structures in all ordered phases of \srmo, \cdmo, and \pbmo\ have been determined. Through comparison with \camo, our results have revealed a universal interplay between magnetic and orbital degrees of freedom in this series of divalent $A$-site quadruple perovskite manganites. In the high temperature magnetically ordered state the periodicities of the magnetic and orbital subsystems lock together via \emph{commensurate} magneto-orbital coupling. On cooling, these periodicities delock, allowing the system to evolve towards a multi-\textbf{k} magnetic ground state in which \emph{incommensurate} magneto-orbital coupling gives rise to a modulation of the spin helicity. 
We have shown that the periodicity of the ground state magnetic structure is controlled by the $A^{2+}$ ionic radius. This in turn determines the temperature at which the magnetic and orbital periodicities de-lock, while the N$\mathrm{\acute{e}}$el temperature remains approximately constant. The universality of our results suggests that these unusual magnetic textures may be observed in other families of manganites with the same 3:1 ratio of 3+ and 4+ ordered $B$-site charge states.

\begin{acknowledgments}
RDJ acknowledges support from a Royal Society University Research Fellowship. AAB was supported by the Japan Society for the Promotion of Science (JSPS) through Grants-in-Aid for Scientific Research (15K14133 and 16H04501) and JSPS Bilateral Open Partnership Joint Research Projects. NT was supported by the JSPS through Grants-in-Aid for Scientific Research (15H05433). PGR acknowledges support from EPSRC, grant number EP/M020517/1, entitled ``Oxford Quantum Materials Platform Grant''.
\end{acknowledgments}

\appendix
\section{\label{appendix}Crystal structure of \cdmo\ below $T_\mathrm{OO}$}
\begin{figure}
\includegraphics[width=8.8cm]{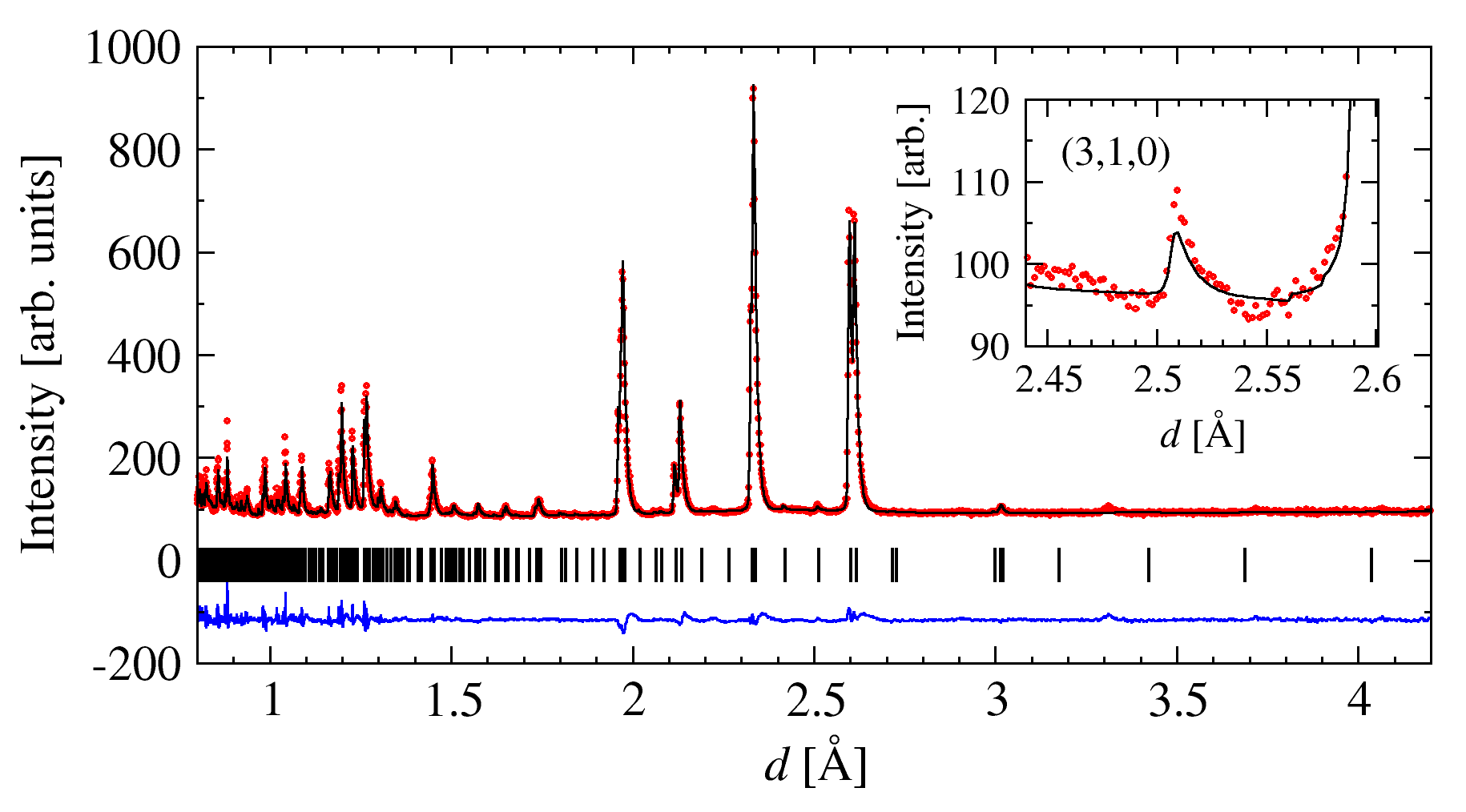}
\caption{\label{CdP3}High resolution neutron powder diffraction data measured in bank 5 (average $2\theta = 152.8^\circ$) of the WISH diffractometer from \cdmo\ at 104 K. Diffraction data are shown as red points, the nuclear peak positions are denoted by the black tick marks, and the fit to the data is shown as a solid black line. A difference pattern ($I_\mathrm{obs}-I_\mathrm{calc}$) is given as a blue solid line at the bottom of the pane. The inset shows a typical $P\bar{3}$ super-structure diffraction peak forbidden in the average $R\bar{3}$ space group.}
\end{figure}

The crystal structure of \cdmo\ was refined against neutron powder diffraction data measured below $T_\mathrm{OO} = 254 \mathrm{K}$, shown in Figure \ref{CdP3}. Atomic displacements that result directly from the orbital order give rise to new diffraction peaks that can be indexed with the commensurate propagation vector $\mathbf{k_s} = (0,0,2)$ (see Figure \ref{CdP3} inset). This propagation vector is equivalent to $(0,0,1)$, however we choose the former to be consistent with the description of magneto-orbital coupling presented in the main text. The propagation vector indicates that translational symmetries [2/3, 1/3, 1/3] and [1/3, 2/3, 2/3] within the $R$-centred conventional unit cell have been broken by the orbital order, and the pattern of atomic displacements can be fully described within space group $P\bar{3}$. The refined atomic fractional coordinates and isotropic atomic displacement parameters are given in Table \ref{Cdnuctab}. Our results are in good agreement with those published in reference \citenum{Guo17} with the exception of the fractional coordinates of atom O1-1 (labelled O1 in Table 2 of ref. \citenum{Guo17}). In the previously published structure \cite{Guo17} this atom appears to deviate greatly from its nominal position in the higher symmetry $R\bar{3}$ structure, and as a result bonding with its neighbours would be unphysical.

\begin{table*}
\caption{\label{Cdnuctab}Crystal structure parameters of \cdmo\ below the orbital ordering phase transition, measured at 104 K, and refined in the commensurate $P\bar{3}$ super-cell. Lattice parameters: $a = 10.4394(2)$, $c = 6.3401(1)$. $R_\mathrm{Bragg}=7.24\%.$}
\begin{ruledtabular}
\begin{tabular}{ccc|ccc|cccc}
\multicolumn{2}{c}{$R\bar{3}$ $(T>T_\mathrm{OO})$} & \multicolumn{2}{c}{$P\bar{3}$ $(T<T_\mathrm{OO})$} \\
Atom & Site & Sym. & Atom & Site & Sym. & $x$ & $y$ & $z$ & U$_\mathrm{iso}$ \\
\hline
\rule{0pt}{4ex} Cd1 & 3a  & $\bar{3}.$ & Cd1-1 & 1a & $\bar{3}..$ & 0         & 0         & 0         & 0.035(4)\\
				    &     &            & Cd1-2 & 2d & $3..$       & 1/3       & 2/3       & 0.659(6)  & 0.035(4)\\
\rule{0pt}{4ex} Mn1 & 9e  & $\bar{1}$  & Mn1-1 & 3e & $\bar{1}$   & 1/2       & 0         & 0         & 0.008(2)\\
				    &     &            & Mn1-2 & 6g & $1$         & 0.183(2)  & 0.337(3)  & 0.328(4)  & 0.008(2)\\
\rule{0pt}{4ex} Mn2 & 9d  & $\bar{1}$  & Mn2-1 & 3f & $\bar{1}$   & 1/2       & 0         & 1/2       & 0.010(2)\\
				    &     &            & Mn2-2 & 6g & $1$         & 0.167(3)  & 0.327(3)  & 0.817(4)  & 0.010(2)\\
\rule{0pt}{4ex} Mn3 & 3b  & $\bar{3}.$ & Mn3-1 & 1b & $\bar{3}..$ & 0         & 0         & 1/2       & 0.010(2)\\
				    &     &            & Mn3-2 & 2d & $3..$       & 1/3       & 2/3       & 0.161(6)  & 0.010(2)\\
\rule{0pt}{4ex} O1  & 18f & $1$        & O1-1  & 6g & $1$         & 0.237(2)  & 0.285(2)  & 0.080(3)  & 0.0132(9)\\
				    &     &            & O1-2  & 6g & $1$         & 0.895(2)  & 0.614(2)  & 0.416(4)  & 0.0132(9)\\
				    &     &            & O1-3  & 6g & $1$         & 0.549(1)  & 0.934(1)  & 0.752(3)  & 0.0132(9)\\
\rule{0pt}{4ex} O2  & 18f & $1$        & O2-1  & 6g & $1$         & 0.346(2)  & 0.522(2)  & 0.352(2)  & 0.0132(9)\\
				    &     &            & O2-2  & 6g & $1$         & 0.007(2)  & 0.860(2)  & 0.684(2)  & 0.0132(9)\\
				    &     &            & O2-3  & 6g & $1$         & 0.327(1)  & 0.815(2)  & 0.010(1)  & 0.0132(9)\\
\end{tabular}
\end{ruledtabular}
\end{table*}

\bibliography{amo}

\end{document}